\documentclass[11pt,a4paper]{article}
\usepackage{jheppub_kim}

\usepackage{pdflscape}
\usepackage{amsmath}
\usepackage{amssymb}
\usepackage{dcolumn}
\usepackage{bm}
\usepackage{color}
\usepackage{epsfig}
\usepackage{amsfonts}
\usepackage{graphicx}
\usepackage{subfigure}
\usepackage{dcolumn}

\newcommand{\be}{\begin{equation}}
\newcommand{\ee}{\end{equation}}
\newcommand{\bea}{\begin{eqnarray}}
\newcommand{\eea}{\end{eqnarray}}

\def\be{\begin{equation}}
\def\ee{\end{equation}}
\def\bea{\begin{eqnarray}}
\def\eea{\end{eqnarray}}

\begin{document}

\title{Thermodynamics in $f(R,\mathcal{L})$ theories: Apparent horizon in the FLRW spacetime}

\author[a]{Behnam Pourhassan,}
\author[b]{Prabir Rudra}

\affiliation[a] {School of Physics, Damghan University, Damghan,
3671641167, Iran.
\\Iran Science Elites Federation, Tehran, Iran.\\
Canadian Quantum Research Center 204-3002 32 Avenue Vernon,
British Columbia V1T 2L7 Canada.} \affiliation[b] {Department of
Mathematics, Asutosh College, Kolkata-700 026, India.}

\emailAdd{b.pourhassan@umz.ac.ir}

\emailAdd{prudra.math@gmail.com, rudra@associates.iucaa.in}

\abstract{In this paper we study the recently proposed
$f(R,\mathcal{L})$ theories from a thermodynamic point of view.
The uniqueness of these theories lies in the fact that the
space-time curvature is coupled to the baryonic matter instead of
exotic matter (in the form of scalar field). We investigate the
viability of these theories from the point of view of the
thermodynamic stability of the models. To be more precise here we
are concerned with the thermodynamics of the apparent horizon of
Friedmann-Lemaitre-Robertson-Walker (FLRW) spacetime in the
background of the $f(R,\mathcal{L})$ theory. We consider several
models of $f(R,\mathcal{L})$ theories where both minimal and
non-minimal coupling has been considered. Various thermodynamic
quantities like entropy, enthalpy, internal energy, Gibbs free
energy, etc. are computed and using their allowed ranges various
model parameters are constrained.}


\maketitle

\section{Introduction}
Late time accelerated expansion \cite{acc1, acc2} of the universe
is the biggest riddle of modern cosmology. Logically thinking,
gravity being attractive in nature will tend to slow down the
expansion of the universe in late times. But the observations are
speaking a completely different story. Indeed the universe has
entered into a phase of accelerated expansion and quite naturally
there has been no satisfactory explanation to this phenomenon till
date. Einstein's theory of general relativity (GR) is totally
inconsistent with this phenomenon and we are left with no choice
other than resorting to modifying the field equations of GR so
that the modified equations can satisfactorily incorporate the
accelerated expansion.

To date all the attempts of modifying the Einstein's field
equations can be broadly classified into two categories. The first
category incorporates exotic nature in the matter content of the
universe, and is termed as \textit{dark energy}. The second
category modifies the gravity component of the equation, thus
bringing about changes in the space-time geometry. This leads to
the concept of \textit{modified gravity theory}. Here we are
interested in this second category which attempts at modifying the
curvature of space-time.

The simplest model of modified gravity is the $\Lambda CDM$ model
where cold dark matter is coupled with the cosmological constant
$\Lambda$. This cosmological constant has an antigravity effect
that drives the accelerated expansion. A special class of models
attempt to modify the gravitational Lagrangian in the
Einstein-Hilbert action by replacing $\mathcal{L}_{GR}=R$ by an
analytic function of the scalar curvature given by
$\mathcal{L}_{f(R)}=f(R)$. This model helps us to explore the
non-linear effects of scalar curvature in the evolution of the
universe by considering arbitrary functions of $R$ in the
gravitational Lagrangian. Extensive reviews in $f(R)$ gravity can
be found in the Refs. \cite{fr1, fr2}. Another class of models
considers non-minimal coupling (NMC) between matter and curvature
\cite{nmc1, nmc2, nmc3, nmc4, nmc5}. These models have been quite
successful at explaining the postinflationary preheating
\cite{inflat0} and cosmological structure formation
\cite{structure1, structure2, structure3}. Further these models
have also been able to successfully mimic dark energy \cite{de1,
de2, de3} and dark matter \cite{dm1, dm2, dm3}.

Most models of NMC have incorporated coupling between curvature
and scalar field \cite{scalcurve1, scalcurve2, scalcurve3,
scalcurve4, scalcurve5, scalcurve6}. But extension of this
coupling to baryonic matter content has been very rare in
literature. Recently, a dynamical system analysis approach was
used to analyze a model that incorporated both $f(R)$ theories and
a NMC with the baryonic matter content \cite{nmc6}. Ref.
\cite{frl} extended this coupling to the baryonic matter content
and studied a more general class of $f(R, \mathcal{L})$ theories
via a dynamical system analysis, where $\mathcal{L}$ represents
the matter Lagrangian. Here we are motivated to study the
thermodynamical aspects of such $f(R, \mathcal{L})$ group of
theories. The motivations to study such theories are quite obvious
and lie in the fact that in these models baryonic matter couples
with space-time curvature. The literature is full with models
involving coupling between space-time curvature and exotic matter
in the form of scalar field \cite{scalcurve1, scalcurve2,
scalcurve3, scalcurve4, scalcurve5, scalcurve6}. But coupling
baryonic matter with curvature seems to be a comparatively alien
topic to cosmologists. But logically this should be the more
realistic scenario because of the non exotic nature of matter. To
be able to describe the universe without resorting to exotic
matter will be a very important step in cosmology. Therein lies
the motivation in studying $f(R,\mathcal{L})$ theories.

It was back in the 1970s when the physicists were first starting
to understand that there was a deep connection between gravity and
thermodynamics. The early form of these ideas was limited to the
study of black hole (BH) thermodynamics. It was found that there
was a link between the horizon area and the entropy of BHs. Since
horizon area is a geometric quantity and entropy is a
thermodynamic quantity, physicists became confident of the deep
underlying connection between the Einstein's field equations and
thermodynamics \cite{bardeen}. Moreover the surface gravity of the
BHs was found to be associated with its temperature and these
quantities followed the first law of thermodynamics (FLT)
\cite{hawking}. Using the fact that entropy is proportional to the
horizon area of BH and the first law of thermodynamics $\delta
Q=Tds$, Jacobson in 1995 derived the Einstein's field equations
\cite{jacobson}. The literature is filled with studies connecting
FLT with Einstein's field equations in modified gravity theories.
Cai and Kim in \cite{cai} derived the Friedmann equations from a
thermodynamic point of view. Akbar in \cite{akbar} discussed the
relation between FLT and Friedmann equations in scalar-tensor
theories and $f(R)$ gravity. Bamba studied the first and the
second laws of thermodynamics in $f(R)$ gravity using the Palatini
formalism \cite{bamba2}. He also studied the thermodynamics of
cosmological horizons in $f(T)$ gravity in \cite{bamba3}. Wu
\emph{et al.} studied the laws of thermodynamics for generalized
$f(R)$ gravity with curvature matter coupling in an universe
described by Friedmann-Lemaitre-Robertson-Walker (FLRW) equations
\cite{wu2}. The laws of thermodynamics at the apparent horizon of
FLRW space-time in background of modified gravity theories is
discussed in \cite{mg2, mg3}. In these studies non-minimal
coupling between matter and space-time geometry has been
considered. Drawing motivations from the above works, here we
intend to study the thermodynamics in $f(R,\mathcal{L})$ gravity
theory. We propose to explore the the effects of both minimal and
non-minimal coupling of matter and geometry on the thermodynamical
aspects of the theory of gravity.

The paper is organized as follows. In Sec.2 we discuss
$f(R,\mathcal{L})$ theories and the basic equations involved.
Section 3 deals with the basic thermodynamical quantities to be
studied for these models. In Sec.4 we discuss these quantities for
general relativity where $R$ is coupled to $\mathcal{L}$
minimally. Section 5 deals with the thermodynamic study of various
non-minimally coupled models. Finally the paper ends with a
discussion and conclusion in Sec.6.

\section{$f(R,\mathcal{L})$ theory of gravity}
The action for the $f(R,\mathcal{L})$ theory \cite{frl} is given
by
\begin{equation}\label{action}
I=\int d^{4}x\sqrt{-g}f(R,\mathcal{L})
\end{equation}
where $f(R,\mathcal{L})$ is a function of both the scalar
curvature $R$ and the matter Lagrangian density $\mathcal{L}$. As
usual $g$ is the determinant of the metric tensor $g_{\mu\nu}$.
This action is a much wider generalization of the Einstein-Hilbert
action than the $f(R)$ theories. Here a non-minimal coupling (NMC)
between the curvature and baryonic matter has been introduced in
the action via the arbitrary function $f(R,\mathcal{L})$. If we
put $f(R,\mathcal{L})=\kappa\left(R-2\Lambda\right)+\mathcal{L}$
(where $\kappa=c^{4}/(16\pi G)$ is a constant) we recover GR with
a cosmological constant $\Lambda$. For any $f(R)$ modification we
can consider $f(R,\mathcal{L})=f(R)+\mathcal{L}$ and for NMC
theories between matter and curvature we can consider
$f(R,\mathcal{L})=f_{1}(R)+f_{2}(R)\mathcal{L}$.

Varying the action with respect to the metric we get the field
equations for the theory as
\begin{equation}\label{fieldeqns}
f^{R}G_{\mu\nu}=\frac{1}{2}g_{\mu\nu}\left(f-f^{R}R\right)+\Delta_{\mu\nu}f^{R}+\frac{1}{2}f^{\mathcal{L}}\left(T_{\mu\nu}-g_{\mu\nu}\mathcal{L}\right)
\end{equation}
where $f^R=\frac{\partial f}{\partial R}$,
$f^\mathcal{L}=\frac{\partial f}{\partial \mathcal{L}}$, and $\Delta_{\mu\nu}=\nabla_{\mu}\nabla_{\nu}-g_{\mu\nu}\Box$. Hence, the
energy momentum tensor is given by,
\begin{equation}\label{em}
T_{\mu\nu}=-\frac{2}{\sqrt{-g}}\frac{\delta(\sqrt{-g}\mathcal{L})}{\delta
g^{\mu\nu}}.
\end{equation}
The conservation equation for this theory turns out to be
\begin{equation}\label{conservation}
\nabla^{\mu}T_{\mu\nu}=\left(g_{\mu\nu}\mathcal{L}-T_{\mu\nu}\right)\left(\frac{f^{R\mathcal{L}}}{f^{\mathcal{L}}}\nabla^{\mu}R+\frac{f^{\mathcal{L}\mathcal{L}}}
{f^{\mathcal{L}}}\nabla^{\mu}\mathcal{L}\right)
\end{equation}
This shows that the conservation equation is no longer covariantly conserved.\\
In order to study the cosmological evolution of the model we can
consider the flat Friedmann-Lemaitre-Robertson-Walker (FLRW) line
element in flat space, which is given by (in unit of light speed),
\begin{equation}\label{flrw}
ds^2=-dt^2+a(t)^2(dx^2+dy^2+dz^2),
\end{equation}
where $a(t)$ is the scale factor or the expansion factor of the
universe \cite{FLRW1, FLRW2}. In that case the Ricci scalar
obtained in terms of scale factor is as follows,
\begin{equation}\label{Ricci}
R=6\left(\frac{\ddot{a}}{a}+\left(\frac{\dot{a}}{a}\right)^{2}\right).
\end{equation}
It is possible to consider FLRW metric in non-flat universe as in
\cite{FLRW2, FLRW3}. We can also consider that matter behaves like
a perfect fluid whose energy-momentum tensor is given by,
\begin{equation}\label{matterem}
T^{\mu\nu}=\left(\rho+p\right)u^{\mu}u^{\nu}+pg^{\mu\nu}.
\end{equation}
It has been found from the matter Lagrangian that \cite{rho1, rho2, rho3},
\begin{equation}\label{Lag}
\mathcal{L}=-\rho.
\end{equation}
Here, $\rho$ and $p$ are the energy density and pressure of the
fluid respectively, and $u^{\mu}$ denotes the four-velocity of the
fluid. In that case, by using the Eq.(\ref{conservation}), one can
obtain the following continuity equation for matter,
\begin{equation}\label{continuity}
\dot{\rho}+3H\left(1+\omega\right)\rho=0
\end{equation}
where $H=\frac{\dot{a}}{a}$ is the Hubble parameter and
$\omega=\frac{p}{\rho}>-\frac{1}{3}$ is the equation of state
(EoS) parameter of the fluid \cite{new} (which we consider as a
constant). From (\ref{continuity}), it is easy to find that
\begin{equation}\label{continuity1}
\rho=\rho_{0}a^{-3(1+\omega)},
\end{equation}
where $\rho_{0}$ is the integration constant. Using the Hubble
expansion parameter in the Eq.(\ref{Ricci}) one can obtain,
\begin{equation}\label{Ricci2}
R=6(\dot{H}+2H^{2}).
\end{equation}
Using the FLRW metric in the field equations one can find that the
first Friedmann equation (the $00-$ component) as,
\begin{equation}\label{frw}
H^{2}=\frac{1}{3f^{R}}\left[\frac{1}{2}f^{R}R-3Hf^{RR}\dot{R}-\frac{1}{2}f-9H^{2}f^{R\mathcal{L}}\left(1+\omega\right)\rho\right],
\end{equation}
Consequently the modified Raychaudhuri equation is given by,
\begin{equation}\label{raychaudhuri}
2\dot{H}+3H^{2}=\frac{1}{2f^{R}}\left[f^{R}R-f-f^{\mathcal{L}}\left(1+\omega\right)\rho-2{\ddot{f}}^{R}-4H{\dot{f}}^{R}\right].
\end{equation}
It is possible to rewrite the above equations as the following forms,
\begin{equation}\label{frw-raychaudhuri}
H^{2}=\frac{8\pi G_{eff}}{3}\rho_{eff},
\end{equation}
and
\begin{equation}\label{frw-raychaudhuri2}
\dot{H}=-4\pi G_{eff}(\rho_{eff}+p_{eff}),
\end{equation}
where $G_{eff}=\frac{G}{f^{R}}$, $\rho_{eff}=\rho+\rho_{m}$ and
$p_{eff}=p+p_{m}$ with $\rho_{m}$ and $p_{m}$ being the density
and pressure contributions from the modified gravity. Using Eqs.
(\ref{frw}), (\ref{raychaudhuri}), (\ref{frw-raychaudhuri}) and
(\ref{frw-raychaudhuri2}) we obtain the density and pressure
contributions from the modified gravity as,
\begin{equation}\label{rho}
\rho_{m}=\frac{1}{8\pi G}\left[\frac{1}{2}f^{R}R
-3Hf^{RR}\dot{R} -\frac{1}{2}f
-9H^{2}f^{RL}(1+\omega)\rho\right]-\rho ,
\end{equation}
and
\begin{equation}\label{pressure}
p_{m}=-\frac{1}{16\pi
G}\left[f^{R}R-f-f^{\mathcal{L}}\left(1+\omega\right)\rho-2{\ddot{f}}^{R}-4H{\dot{f}}^{R}\right]-\omega
\rho,
\end{equation}
The effective equation of state (EoS) can be given by,
\begin{equation}\label{EoS}
\omega_{eff}=\frac{p+p_{m}}{\rho+\rho_{mo}}=\frac{p_{eff}}{\rho_{eff}}=-\frac{f^{R}R-f-f^{\mathcal{L}}\left(1+\omega\right)\rho-2{\ddot{f}}^{R}-4H{\dot{f}}^{R}}{f^{R}R
-6Hf^{RR}\dot{R}-f-18H^{2}f^{RL}(1+\omega)\rho},
\end{equation}
Now, combining the Eqs.(\ref{Lag}) and (\ref{frw-raychaudhuri}) we
get the following relation for the matter Lagrangian,
\begin{equation}\label{Lag0}
{\mathcal{L}}=\rho_{m}-\frac{3H^{2} f^{R}}{8\pi G},
\end{equation}
Here, we are interested in studying the thermodynamical properties
of the model, where matter Lagrangian couple to the curvature
scalar. In the next section we study the basic thermodynamical
parameters which we will use in our investigation.

\section{Thermodynamics in $f(R,\mathcal{L})$ gravity}
In this section we will compute the basic thermodynamic quantities
in a model independent way. Using the allowable region for these
quantities we can check the stability and viability of the model.
We may also constrain some model parameters using these relations.
The apparent horizon radius for FLRW universe is given by,
\begin{equation}\label{ah}
r_{A}=H^{-1}.
\end{equation}
Hence, the Eq.(\ref{Ricci2}) yields the following expression,
\begin{equation}\label{Ricci3}
R=\frac{6}{r_{A}^{2}}(2-\dot{r}_{A}).
\end{equation}
Therefore, one can obtain,
\begin{equation}\label{Ricci-t}
\dot{R}=\frac{6(2{\dot{r}_{A}}^{2}-r_{A}{\ddot{r}}_{A}-4\dot{r}_{A})}{r_{A}^{3}}.
\end{equation}
and
\begin{equation}\label{Ricci-r}
R^{\prime}=-\frac{6\left(\frac{r_{A}{\ddot{r}}_{A}}{\dot{r}_{A}}+2(2-\dot{r}_{A})\right)}{r_{A}^{3}},
\end{equation}
where dot denotes derivative with respect to time  and prime
denotes derivative with respect to $r_{A}$. The associated
temperature of the apparent horizon is given by \cite{cai},
\begin{equation}\label{T}
T=\frac{1}{4\pi}(2H+\frac{\dot{H}}{H})=\frac{|1-\frac{{\dot{r}}_{A}}{2}|}{2\pi r_{A}},
\end{equation}
while the entropy is,
\begin{equation}\label{S}
S=\frac{Af^{R}}{4G},
\end{equation}
where $A=4\pi r_{A}^{2}$ is apparent horizon area with volume
$V=\frac{4}{3}\pi r_{A}^{3}$. Hence, all thermodynamic variables
could be expressed in terms of apparent horizon area and its
derivatives. Temperature variation with respect to horizon radius
yields,
\begin{equation}\label{dT}
\frac{dT}{dr_{A}}=\frac{1}{4\pi r_{A}^{2}}\left|\frac{r_{A}{\ddot{r}}_{A}}{{\dot{r}}_{A}}-{\dot{r}}_{A}+2\right|,
\end{equation}
while entropy variation yields,
\begin{equation}\label{dS}
\frac{dS}{dr_{A}}=\frac{\pi}{G}r_{A}\left(2f^{R}+r_{A}\frac{df^{R}}{dr_{A}}\right).
\end{equation}
The specific heat, which is an important thermodynamic quantity to
study model stability, is given by,
\begin{equation}\label{C}
C=T\frac{dS}{dT}=\frac{\pi}{G}\frac{(2-{\dot{r}}_{A})r_{A}^{2}{\dot{r}}_{A}(2f+r_{A}\frac{df^{R}}{dr_{A}})}{\left|r{\ddot{r}}_{A}-{\dot{r}}_{A}^{2}+2{\dot{r}}_{A}\right|}.
\end{equation}
If the sign of specific heat is positive then the cosmological
model is stable, and it helps us to realize the allowed region of
the apparent horizon where the model is stable. By using the
Eq.(\ref{S}), one can obtain,
\begin{equation}\label{dS}
dS=\frac{Adf^{R}+f^{R}dA}{4G},
\end{equation}
where $dA=8\pi r_{A}dr_{A}$. In this case the internal energy
(total energy) could be expressed as \cite{new},
\begin{equation}\label{E}
\hat{E}=\frac{3f^{R}H^{2}}{8\pi G}V=V\rho_{eff}
\end{equation}
Therefore,
\begin{equation}\label{E2}
d\hat{E}=Vd\rho_{eff}+\rho_{eff} dV.
\end{equation}
Hence, the first law of thermodynamics is reduced to the following
expression \cite{new},
\begin{equation}\label{first-law}
Td\hat{S}=d\hat{E}-\hat{W}dV,
\end{equation}
where
\begin{equation}\label{first-law-def1}
d\hat{S}=dS+\frac{(1-2\pi r_{A}T)r_{A}}{2GT}df^{R},
\end{equation}
and
\begin{equation}\label{first-law-def2}
\hat{W}=\frac{\rho_{eff}-p_{eff}}{2}.
\end{equation}
Following \cite{wu2}, it is quite straightforward to show that we
can get the FLRW Eqn.(\ref{frw}) starting from the above mentioned
first law of thermodynamics (Eqn.(\ref{first-law})). We can
propose the modified or corrected entropy for the modified gravity
theory as,
\begin{equation}\label{entropy-corrected}
\hat{S}=S-\frac{3}{G}\int{\frac{(1-2\pi r_{A}T)\left(2{\dot{r}}_{A}(2-{\dot{r}}_{A})+r_{A}{\ddot{r}}_{A}\right)f^{RR}}{Tr_{A}^{2}{\dot{r}}_{A}}dr}.
\end{equation}
Having time-dependent $r_{A}$, we can calculate modified entropy
and other thermodynamic quantities. Therefore, by using the
Eq.(\ref{C}), the specific heat at constant volume (modified
specific heat) is given by the following relation,
\begin{equation}\label{CV}
C_{V}=\left(T\frac{d\hat{S}}{dT}\right)_{V}=\left(\frac{d\hat{E}}{dT}\right)_{V}.
\end{equation}
Then, the specific enthalpy in unit mass is obtained via,
\begin{equation}\label{H}
h=\hat{E}+p_{eff}V,
\end{equation}
which is in turn used to obtain specific Gibbs free energy,
\begin{equation}\label{G}
g=h-T\hat{S},
\end{equation}
and the Helmholtz free energy
\begin{equation}\label{F}
F=\hat{E}-T\hat{S}.
\end{equation}
We will study the thermodynamics of this theory in detail for some
specified models as examples in the sections to follow. We will
basically examine three basic thermodynamic requirements in the
chosen $f(R, \mathcal{L})$ models. First of all we would like to
satisfy the first and second laws of thermodynamics, where the
entropy is an increasing function of time. The second is
cosmological point of view where the temperature of the universe
is a decreasing function of time (with positive value). Finally we
check thermodynamical stability of model by analyzing specific
heat and other thermodynamic potentials. In summary, to have a
well defined model, we should find entropy as an increasing
function, $T\geq0$ (and decreasing with time), $C_{V}\geq 0$, $g$
as a decreasing function and $F$ should have a minimum.

\section{General relativity}
\begin{figure}[h!]
 \begin{center}$
 \begin{array}{cccc}
\includegraphics[width=60 mm]{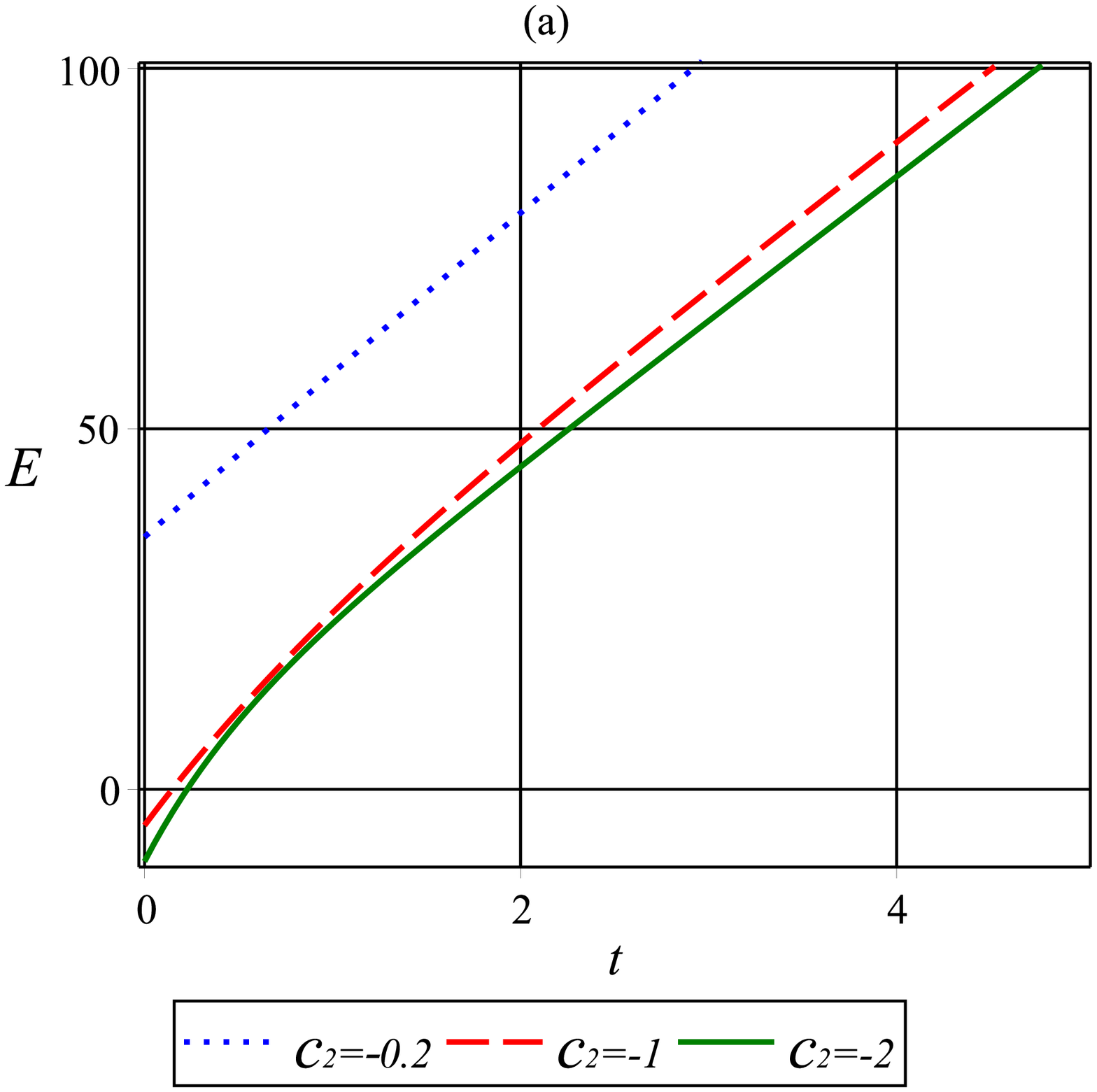}~~~~~~~~\includegraphics[width=60 mm]{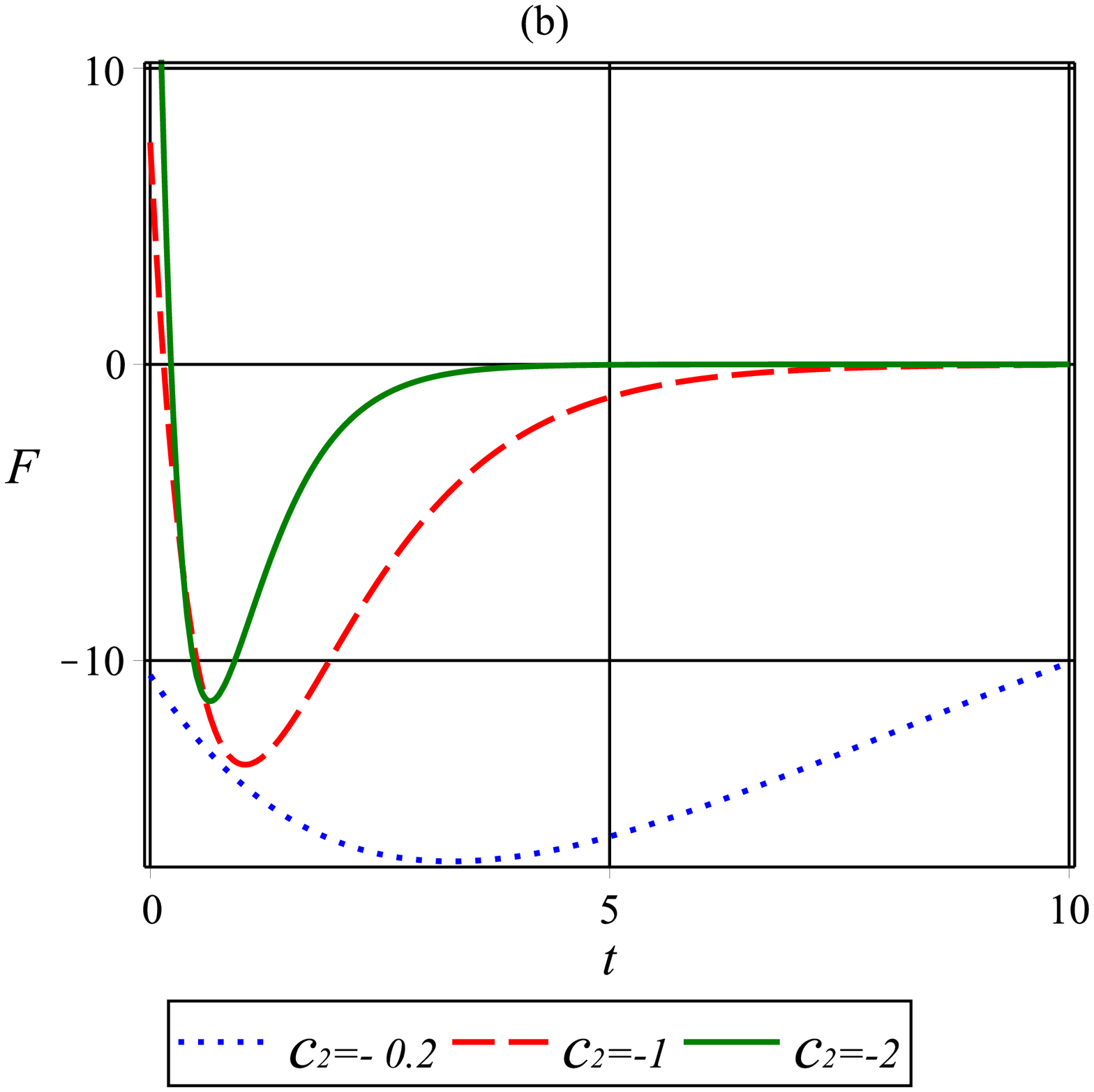}
 \end{array}$
 \end{center}
\caption{Thermodynamic potentials in terms of time $t$ in general
relativity for $\kappa_{1}=1$ and $4\pi G=1$.
(a) shows internal energy $\hat{E}$, (b) shows
Helmholtz free energy.}
 \label{fig2}
\end{figure}

Here, we consider the simplest model of the $f(R,\mathcal{L})$
theories. We minimally couple $R$ with $\mathcal{L}$ in such a way
that we eventually construct general relativity with a
cosmological constant. A study of this model will simply help us
to check the sanity of the model from the thermodynamical point of
view. The model is given by,
\begin{equation}\label{GR}
f(R,\mathcal{L})=\kappa_{1}\left(R-2\Lambda\right)+\mathcal{L},
\end{equation}
where $\kappa_{1}$ is a constant. Hence, $f^{R}=\kappa_{1}$,
$f^{\mathcal{L}}=1$ are the only non-zero derivatives. In that
case, by using the relations (\ref{Lag0}) and (\ref{ah}) one can
find,
\begin{equation}\label{Lag1}
{\mathcal{L}}=2\kappa_{1}(\Lambda-\frac{3}{r_{A}^{2}}).
\end{equation}
Hence, the Eq.(\ref{S}) yields the following entropy,
\begin{equation}\label{S1}
S=\frac{\pi\kappa_{1}}{G}r_{A}^{2}=\hat{S},
\end{equation}
where in the last equality we used the equation
(\ref{entropy-corrected}). Therefore using the energy density
(\ref{rho}) we get the following relation for effective energy
density,
\begin{equation}\label{rho1}
\rho_{eff}=\frac{3\kappa_{1}}{8\pi G}\frac{1}{r_{A}^{2}}.
\end{equation}
Also using the pressure given by Eq.(\ref{pressure}) we get the
following relation,
\begin{equation}\label{pressure1}
p_{eff}=-\frac{1}{16\pi G}\left[\frac{3}{r_{A}^{2}}-(1+\omega)\rho\right],
\end{equation}
therefore effective equation of state obtained from the
Eq.(\ref{EoS}) is as follows,
\begin{equation}\label{EoS1}
\omega_{eff}=-\frac{1}{6\kappa_{1}}[3-(1+\omega)\rho r_{A}^{2}].
\end{equation}
Now, from the Eq.(\ref{first-law-def2}) we obtain,
\begin{equation}\label{first-law-def2-1}
\hat{W}=\frac{3\kappa_{1}}{8\pi G}\frac{1-\omega_{eff}}{2r_{A}^{2}}.
\end{equation}
Hence, the internal energy from the Eq.(\ref{E}) is obtained as,
\begin{equation}\label{E1}
\hat{E}=\frac{\kappa_{1}r_{A}}{2 G}.
\end{equation}
Satisfying the first law of thermodynamics gives us the following
relation,
\begin{equation}\label{first-law1}
r_{A}=4t-3\int\omega_{eff}dt.
\end{equation}
In the case of $\omega_{eff}=-1$ we can see linear behavior with
time as $r_{A}=7t+r_{0}$, where $r_{0}$ is the integration
constant which we consider as a positive parameter. The initial
value of the apparent horizon radius can be fixed by using the
positivity of specific heat which will be
discussed below.\\
By using the Eq.(\ref{CV}) one can obtain,
\begin{equation}\label{CV1}
C_{V}=\frac{2\pi \kappa_{1}}{G}\left(\frac{r_{A}^{2}{\dot{r}}_{A}}{r_{A}{\ddot{r}}_{A}-{{\dot{r}}_{A}}^{2}+2{\dot{r}}_{A}}\right).
\end{equation}
In order to have positive $C_{V}$ both numerator and denominator
should be positive. In the case of $\omega_{eff}<-\frac{4}{3}$, numerator is completely positive.\\
From the denominator we have,
\begin{equation}\label{denominator1}
r_{A}{\ddot{r}}_{A}-{{\dot{r}}_{A}}^{2}+2{\dot{r}}_{A}>0.
\end{equation}
We can solve above equation analytically to obtain,
\begin{equation}\label{sol1}
r_{A}>\frac{c_{1}e^{c_{2}t}-2}{c_{2}},
\end{equation}
where $c_{1}$ and $c_{2}$ are integration constants.
We can write the following ansatz for the apparent horizon radius in agreement with (\ref{sol1}),
\begin{equation}\label{sol11}
r_{A}=c_{3}t+\frac{c_{1}}{c_{2}}e^{c_{2}t}-\frac{2}{c_{2}}.
\end{equation}
Comparing (\ref{first-law1}) and (\ref{sol11}) we get $c_{3}=4$,
$c_{2}<0$ and $c_{1}>0$ so that,
\begin{equation}\label{sol111}
r_{A}=4t-3e^{c_{2}t}-\frac{2}{c_{2}}.
\end{equation}
It is indeed the apparent horizon radius where the model is stable
and the first and second laws of thermodynamics are satisfied as
the entropy is an increasing function of time with positive value.
It means that,
\begin{equation}\label{effective-omega1}
\omega_{eff}=c_{2}e^{c_{2}t}.
\end{equation}
Also, we find the temperature to be a decreasing function of time,
making the model cosmologically viable.\\
Finally it is quite straightforward to investigate the nature of
thermodynamic quantities like internal energy [Fig. \ref{fig2}
(a)]. It is clear that internal energy is an increasing function
of time, which is caused by the accelerated expansion of universe.
From the Eq.(\ref{F}) one can obtain Helmholtz free energy as,
\begin{equation}\label{F1}
F=-\frac{3\kappa_{1}}{4G}e^{c_{2}t}(3c_{2}e^{c_{2}t}-4c_{2}t+2).
\end{equation}
In the Fig. \ref{fig2} (b) we can see behavior of Helmholtz free
energy in terms of time. It is seen that the Helmholtz free energy
has a minimum. These indicate the stability of the model which is
obtained by using the relation (\ref{sol11}). Hence, by using the
Eq.(\ref{ah}) we can obtain Hubble expansion parameter which is
decreasing function of time.

\section{Nonminimally coupled (NMC) Theories}
Although minimally coupled theories are widely found in literature
because of their computational convenience, it is believed that at
high space-time curvatures non-minimal coupling will have a big
role to play, specially in quantum field theory. In fact
non-minimal coupling is introduced by quantum fluctuations and so
they are almost non-existent in classical action \cite{linde1}.
The coupling is actually required if a scalar field theory is to
be renormalized in a classical gravitational background
\cite{freed1}. Here we proceed to study some models where
non-minimal coupling is in action. The results obtained from such
models will be more generic and realistic in nature and hence very
important for the present study.

\subsection{Exponential model}
An interesting model involving NMC coupling between matter and
curvature is the exponential model \cite{expo, frl}, given by
\begin{equation}\label{ex}
f(R,\mathcal{L})=M^{4}e^{(\frac{R}{6H_{0}^{2}}+\frac{\mathcal{L}}{6\kappa_{2} H_{0}^{2}})},
\end{equation}
where $\kappa_{2}$ is a constant, $M$ denotes mass scale, while
$H_{0}$ is related to the expansion rate (current value of the
Hubble parameter which according to the latest observational data
is $H_{0}\approx67$ \cite{planck}). Here, we can see that on
expanding the exponential function in series we will get terms of
the form $\xi f(R) g(\mathcal{L})$ where $\xi$ is a constant. So
here the coupling between $R$ and $\mathcal{L}$ is non-minimal in
nature. It is to be noted that this model does not simplify to
general relativity with a cosmological constant for small $R$ and
$\mathcal{L}$. In this case, by using the relations (\ref{Lag0})
and (\ref{ah}) one can find,
\begin{equation}\label{Lag2}
{\mathcal{L}}=\frac{\kappa_{2}l_{0}^{2}r_{A}^{2}}{9(1+\omega)}
\left(\frac{1}{2}+\frac{18(2{{\dot{r}}_{A}}^{2}-r_{A}{\ddot{r}}_{A}-4{\dot{r}}_{A})}{l_{0}^{2}r_{A}^{4}}+\frac{3}{l_{0}r_{A}^{2}}-\frac{3(2-{\dot{r}}_{A})}{r_{A}^{2}}\right),
\end{equation}
where we used $l_{0}\equiv6H_{0}^{2}$.\\
In that case one can obtain,
\begin{eqnarray}\label{ex-der2}
{\dot{f}}^R&=&\frac{f}{l_{0}^{2}}(\dot{R}+\frac{\dot{\mathcal{L}}}{\kappa_{2}}),\nonumber\\
{\ddot{f}}^{R}&=&\frac{f}{l_{0}^{2}}(\ddot{R}+\frac{\ddot{\mathcal{L}}}{\kappa_{2}})+\frac{f}{l_{0}^{3}}(\dot{R}+\frac{\dot{\mathcal{L}}}{\kappa_{2}})^{2},
\end{eqnarray}
Here $\dot{R}$ is given by the Eq.(\ref{Ricci-t}) and
\begin{equation}\label{L-t}
\dot{\mathcal{L}}=\frac{\kappa_{2}l_{0}^{2}}{9(1+\omega)}\left[r_{A}{\dot{r}}_{A}+{\ddot{r}}_{A}+\frac{18}{l_{0}^{2}}
\frac{7r_{A}{\dot{r}}_{A}{\ddot{r}}_{A}-r_{A}^{2}{\dddot{r}}_{A}-4r_{A}{\ddot{r}}_{A}-4{\dot{r}}_{A}^{3}+16{\dot{r}}_{A}^{2}}{r_{A}^{5}}\right].
\end{equation}
Therefore, we get the following expression for entropy,
\begin{equation}\label{S2}
S\approx\frac{\pi M^{4}}{9G(1+\omega)}\left(\frac{1}{2}-\frac{3(2-\dot{r}_{A})}{r_{A}^{2}}+\frac{3}{l_{0}r_{A}^{2}}\right)r_{A}^{4},
\end{equation}
where we neglected ${\mathcal{O}}(\frac{1}{l_{0}^{2}})$.\\
Combining Eqs.(\ref{Lag}), (\ref{rho}) and (\ref{Lag2}) give us
the effective energy density as follow,
\begin{equation}\label{rho2}
\rho_{eff}=\frac{3}{l_{0}r_{A}^{2}}.
\end{equation}
At leading order it is indeed the Eq.(\ref{rho1}) where replacement $\kappa_{1} l_{0}\rightarrow 8\pi G$ is performed.\\
Then, we can obtain effective pressure as,
\begin{equation}\label{pressure2}
p_{eff}\approx\frac{M^{4}}{16\pi G}\left[1-\left(\frac{1}{2}-\frac{3(2-\dot{r}_{A})}{r_{A}^{2}}+\frac{3}{l_{0}r_{A}^{2}}\right)
\left(\frac{3(2+\omega)(2-\dot{r}_{A})}{2(1+\omega)}+\frac{l_{0}\omega r_{A}^{2}}{9(1+\omega)}+\frac{l_{0}^{2}r_{A}^{4}}{81(1+\omega)}\right)\right],
\end{equation}
where, again, we neglected ${\mathcal{O}}(\frac{1}{l_{0}^{2}})$.\\
Therefore, by using the Eq.(\ref{first-law-def2}) we can obtain,
\begin{equation}\label{first-law-def2-2}
\hat{W}=\frac{M^{4}}{32\pi G}\left[\frac{48\pi G}{l_{0}r_{A}^{2}M^{4}}-1+\left(\frac{1}{2}-\frac{3(2-\dot{r}_{A})}{r_{A}^{2}}+\frac{3}{l_{0}r_{A}^{2}}\right)
\left(\frac{3(2+\omega)(2-\dot{r}_{A})}{2(1+\omega)}+\frac{l_{0}\omega r_{A}^{2}}{9(1+\omega)}+\frac{l_{0}^{2}r_{A}^{4}}{81(1+\omega)}\right)\right].
\end{equation}
Hence we can write,
\begin{equation}\label{EX}
\hat{E}=\frac{4\pi r_{A}}{l_{0}}.
\end{equation}
Because $f^{RR}\propto\frac{1}{l_{0}^{2}}$ hence $\hat{S}\approx S$, and the first law of thermodynamics read as,
\begin{equation}\label{first law 2}
TdS=\frac{4\pi}{l_{0}}dr_{A}-4\pi\hat{W}r_{A}^{2}dr_{A}.
\end{equation}
Putting all the above results in the Eq.(\ref{first-law}) and
solving the resulting differential equation we see that the
apparent horizon has the following form,
\begin{equation}\label{sol3}
r_{A}=r_{0}\tanh{(\omega_{0}t)}.
\end{equation}
where $r_{0}$ and $\omega_{0}$ are arbitrary parameters.

\begin{figure}[h!]
 \begin{center}$
 \begin{array}{cccc}
\includegraphics[width=50 mm]{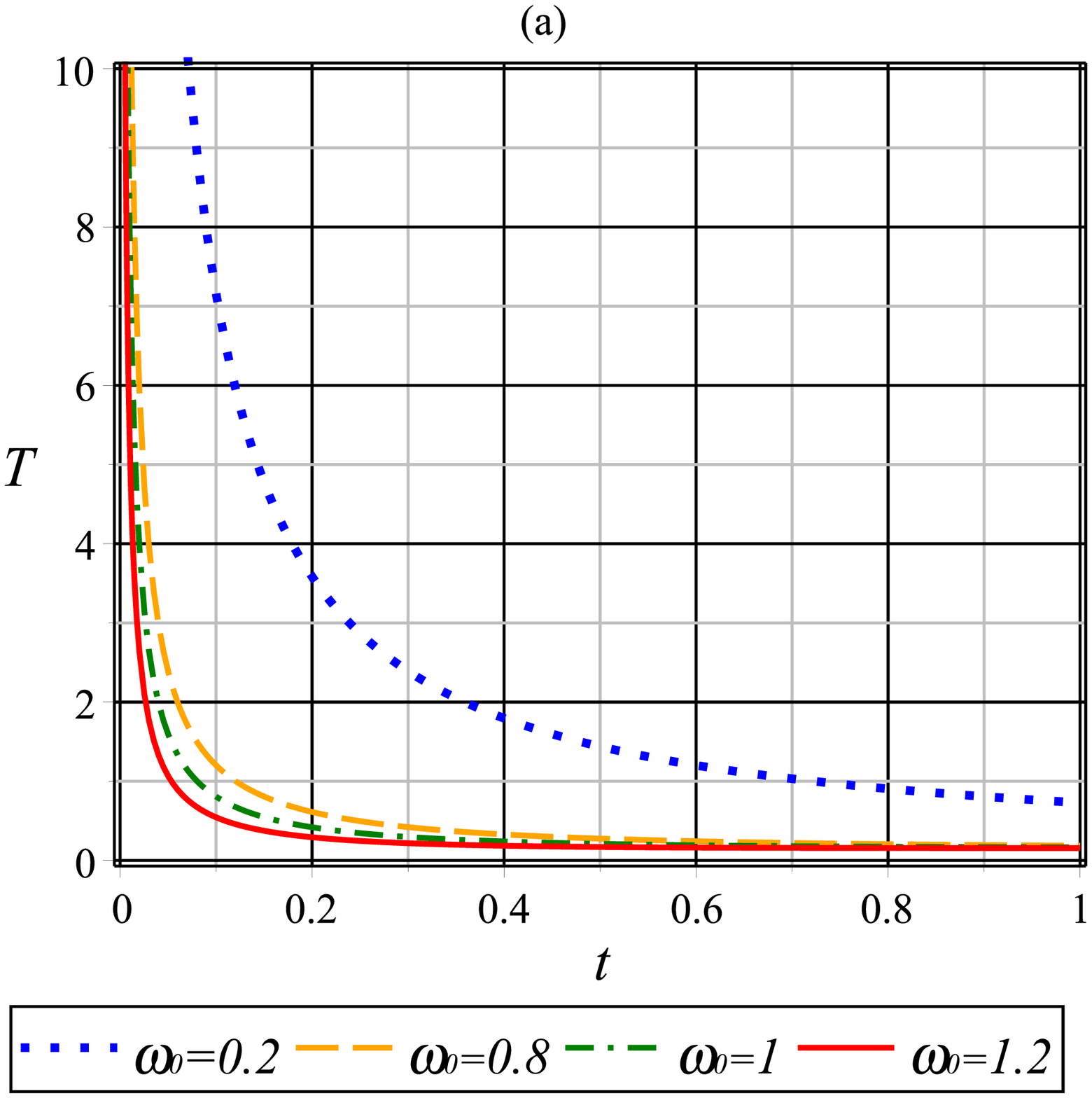}\includegraphics[width=50 mm]{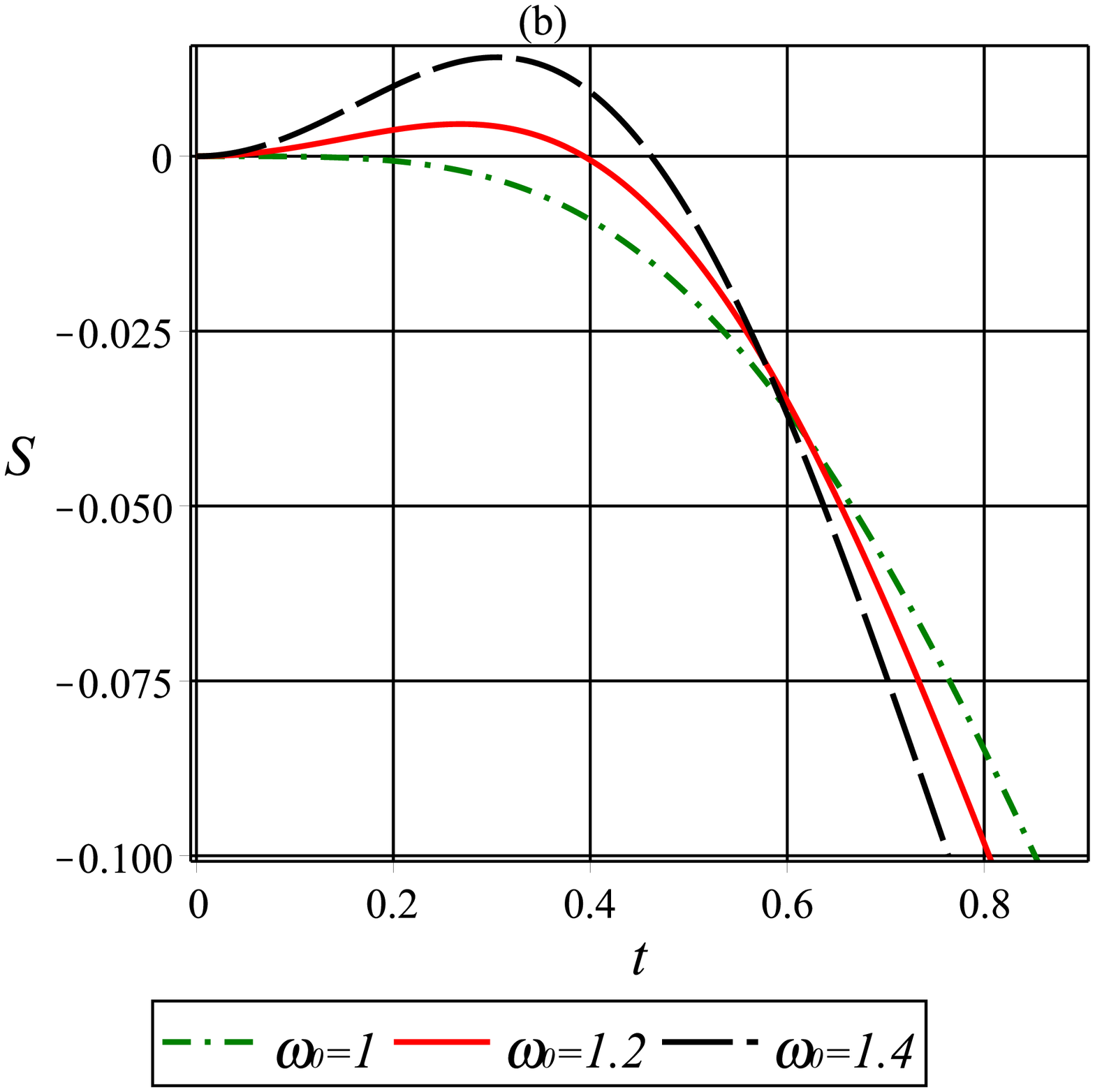}\includegraphics[width=50 mm]{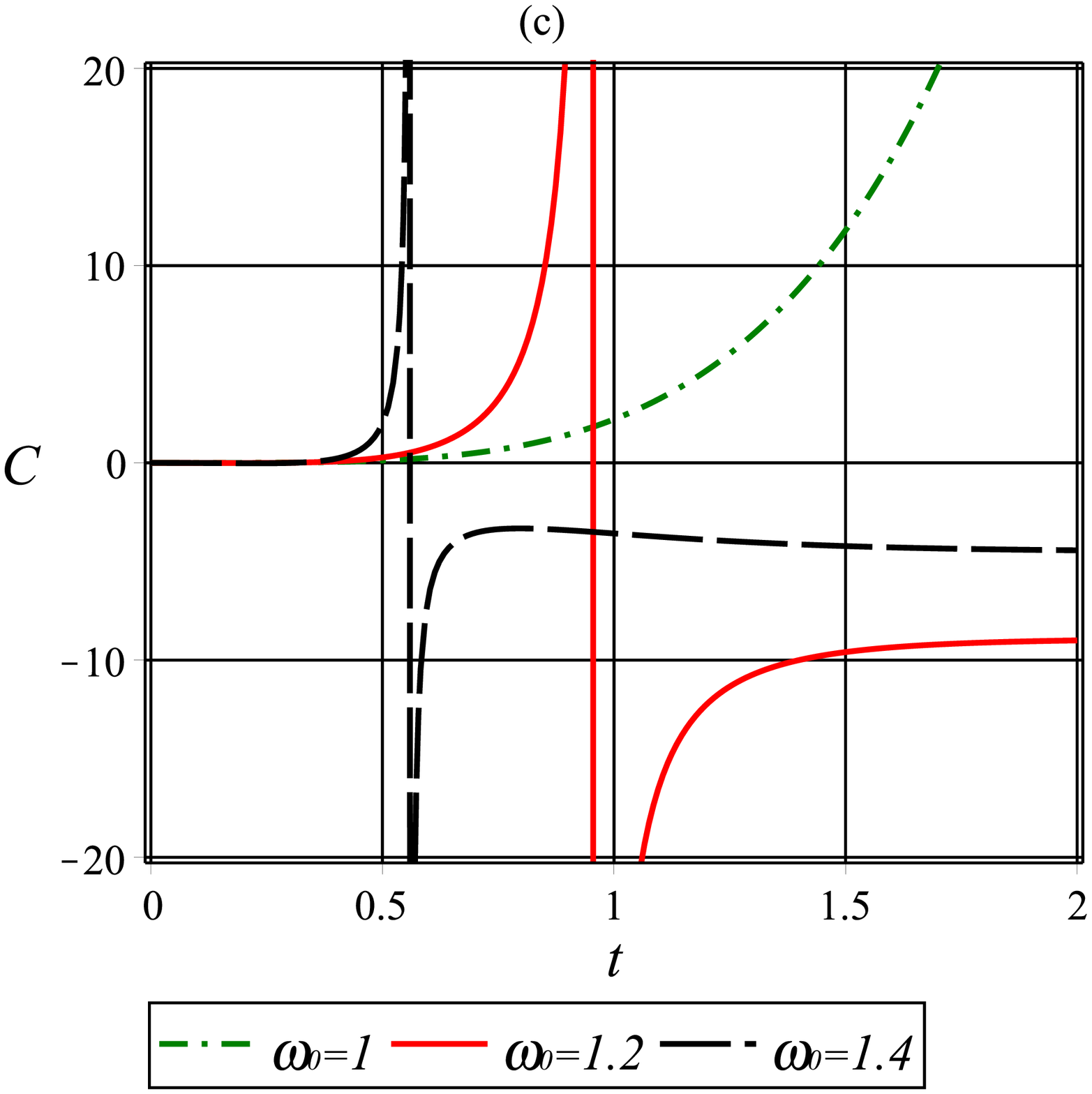}
 \end{array}$
 \end{center}
\caption{Typical behavior of thermodynamics quantities for unit values of parameters.}
 \label{fig3}
\end{figure}

It yields to a negative pressure which is cause of accelerating
expansion. Thermodynamics of this model strongly depends on values
of $r_{0}$ and $\omega_{0}$. In order to have a well-defined model
we should have positive temperature which decreases with time.
Figure \ref{fig3} (a)  shows variation of temperature by time for
specific choice of model parameters. For example in the case of
$r_{0}=1$ and $\omega_{0}=1.2$ we have positive temperature which
is decreasing function of time. In Fig. \ref{fig3} (b) We see that
lower values of $\omega_{0}$ generally yields negative entropy
throughout. For the case of $\omega_{0}=1.2$ we have positive
entropy in the beginning where it is an increasing function of
time. But after reaching a maximum it flips to the negative
region. Some instabilities and phase transitions are illustrated
in Fig. \ref{fig3} (c). For the case of $\omega_{0}=1.2$ we see
asymptotic behavior which shows a phase transition at the late
time. Other thermodynamics potentials like internal energy,
Helmholtz free energy, Gibbs free energy and enthalpy all are
increasing function of time with general behavior like,
\begin{equation}\label{EX2}
\hat{E}=\frac{4\pi}{l_{0}} r_{0}\tanh{(\omega_{0}t)}.
\end{equation}

\subsection{Power law model}
\begin{figure}[h!]
 \begin{center}$
 \begin{array}{cccc}
\includegraphics[width=60 mm]{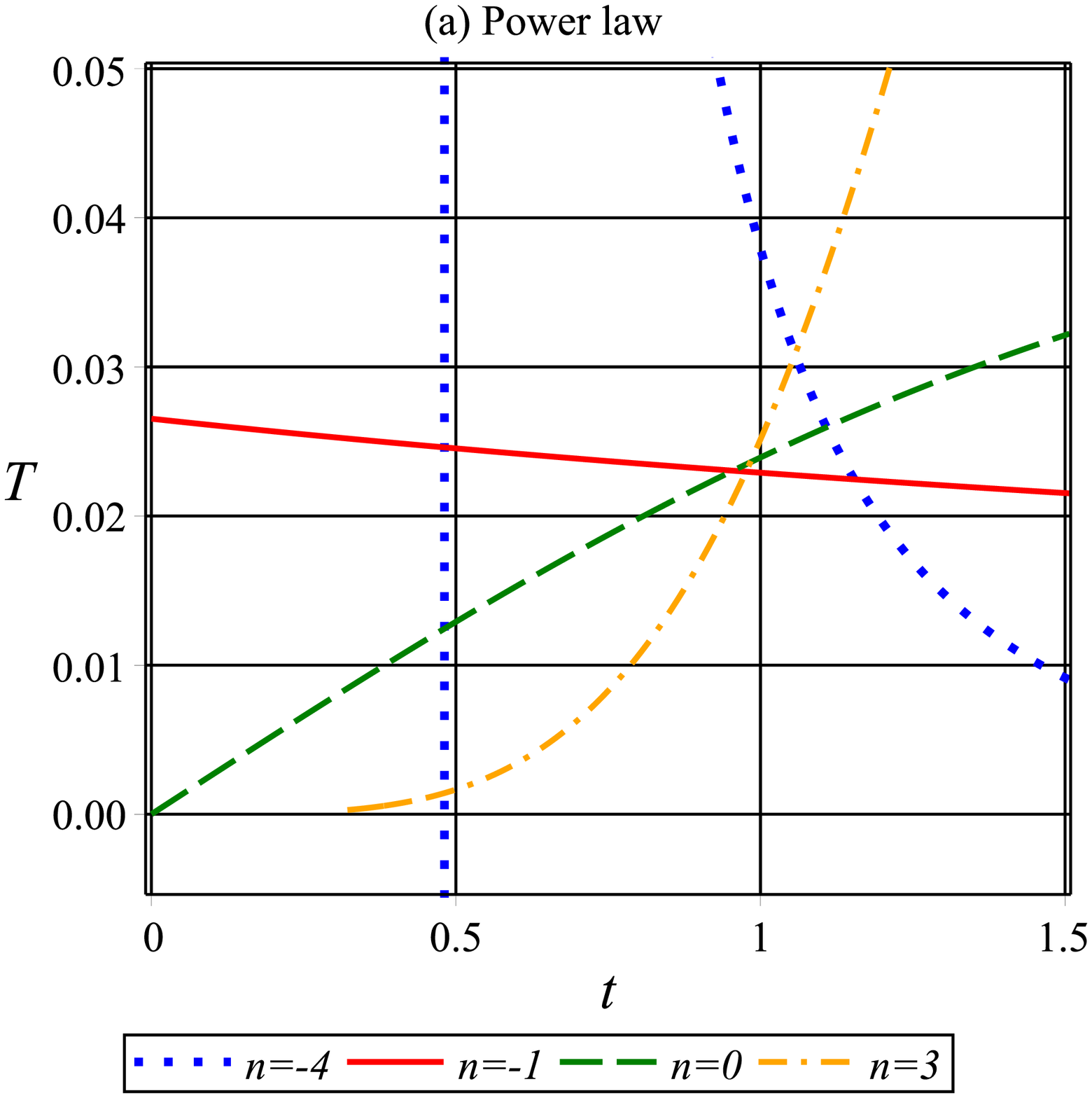}~~~~~~~~\includegraphics[width=60 mm]{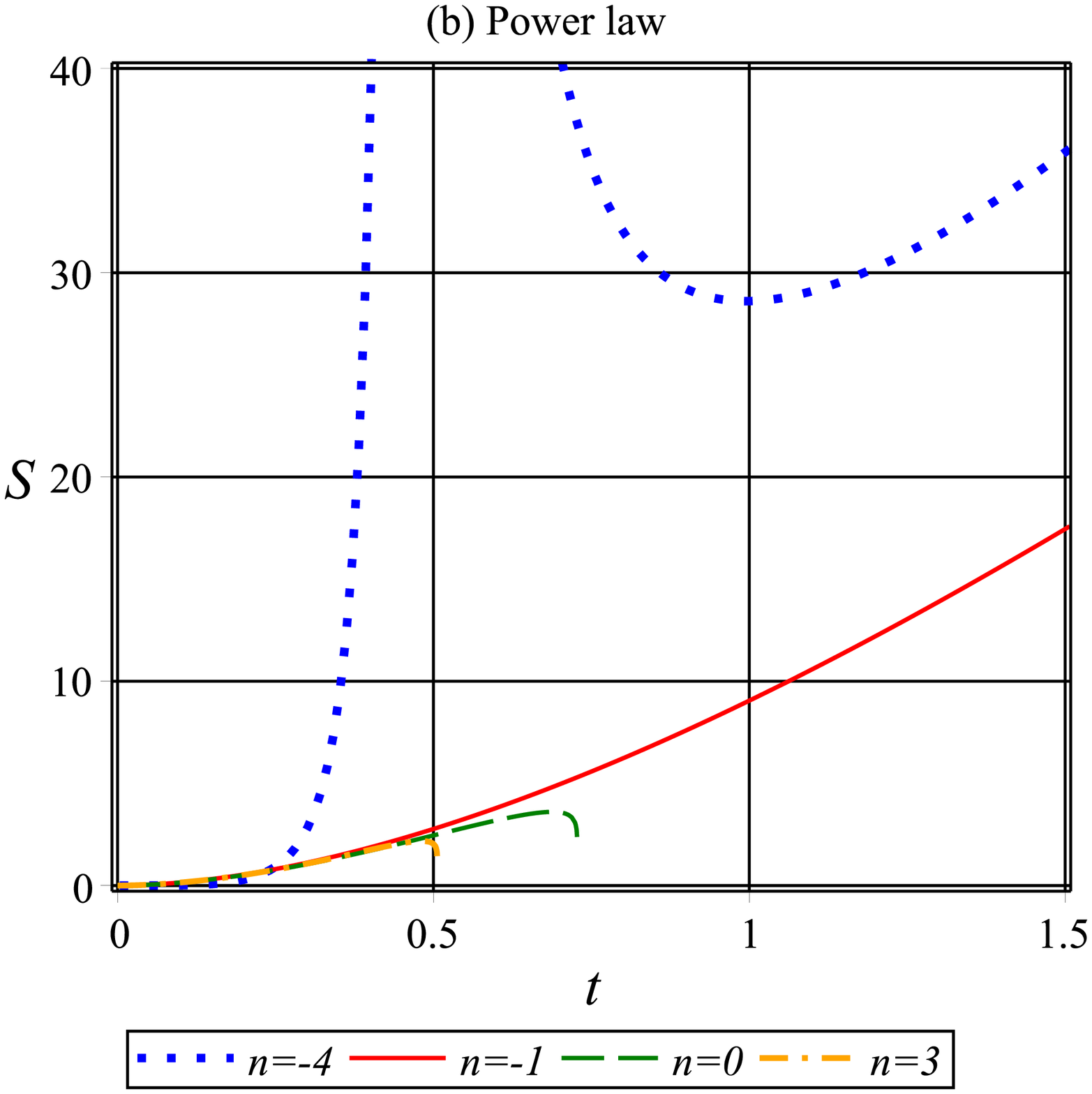}
 \end{array}$
 \end{center}
\caption{Typical behavior of temperature and entropy in terms of
time for $R_{0}=1$. (a) temperature; (b) entropy for
$M=\kappa=G=1$, $\kappa_{3}=0.2$, $\omega=-0.2$ and
$\varepsilon=0.1$.}
 \label{fig4}
\end{figure}

Here, we consider the model
\cite{frl},
\begin{equation}\label{power}
f(R,\mathcal{L})=(\kappa M^{2})^{-\varepsilon}(\kappa_{3} R+\mathcal{L})^{1+\varepsilon},
\end{equation}
where $\kappa_{3}$ is a constant, $M$ denotes mass scale, while
$\varepsilon$ is infinitesimal correction parameter. At the
$\varepsilon\rightarrow0$ limit we recover GR with $\Lambda=0$.
Here for $\epsilon\neq 0$ we get non-minimal terms on power
expansion.\\
Using the relations (\ref{Lag}), (\ref{rho}) and  (\ref{Lag0}) one can obtain,
\begin{equation}\label{Lag3}
{\mathcal{L}}=\frac{\kappa_{3}}{2r_{A}^{2}}\left[(\varepsilon-1)Rr_{A}^{2}+18\varepsilon(\varepsilon+1)\omega+6\varepsilon(3\varepsilon+2)-6
+\sqrt{(\varepsilon+1)\Xi}\right],
\end{equation}
where
\begin{eqnarray}\label{Lag33}
\Xi&\equiv&324(1+\omega)^{2}\varepsilon^{3}+36(1+\omega)(Rr_{A}^{2}+9\omega+3)\varepsilon^2\nonumber\\
&+&(R^{2}r_{A}^{4}-24\dot{R}r_{A}^{3}-(3\omega+4)12Rr_{A}^{2}-216\omega-180)\varepsilon+(Rr_{A}^{2}-6)^2.
\end{eqnarray}
We can neglect the higher power of $\varepsilon$ for simplicity.\\
Power law model of cosmic expansion suggests the following
expression for Ricci scalar,
\begin{equation}\label{Ricci33}
R=R_{0}t^{n}.
\end{equation}
For the $f(R)$ gravity the most favored value of $n$ is found to
be $n=-2$ \cite{0906.3860}, but in $f(R, {\mathcal{L}})$ we can
fix $n$ by using thermodynamic requirements. This choice
simplifies our calculations to obtain apparent horizon radius.
Combination of (\ref{Ricci3}) and (\ref{Ricci33}) give us apparent
horizon as follows,
\begin{equation}\label{sol3}
r_{A}=-2\frac{\sqrt{-3R_{0}}}{R_{0}\sqrt{t^{n}}}\left[\frac{J_{m}(X_{p})+c_{1}Y_{m}(X_{p})}{J_{l}(X_{p})+c_{1}Y_{l}(X_{p})}\right],
\end{equation}
where $c_{1}$ is the integration constant, $m=\frac{1}{n+2}$,
$l=-\frac{n+1}{n+2}$ and
\begin{equation}\label{Bessel3}
X_p=\frac{2\sqrt{-3R_{0}}}{3(n+2)}t^{1+\frac{n}{2}}.
\end{equation}
In the above expressions $J_{m}(X_{p})$ is Bessel function of the
first kind while $Y_{m}(X_{p})$ is Bessel function of the second
kind. In order to have real valued apparent horizon and other
thermodynamic quantities, we should choose $c_{1}=0$ and we can
analyze for both positive and negative Ricci scalar. Hence, we
consider,
\begin{equation}\label{sol3-2}
r_{A}=-2\frac{\sqrt{-3R_{0}}}{R_{0}\sqrt{t^{n}}}\left[\frac{J_{m}(X_{p})}{J_{l}(X_{p})}\right],
\end{equation}
By analyzing the temperature we can fix $n$ in the relation
(\ref{Ricci33}) and hence we can study all cosmological
parameters. Actually we use two requirements: the temperature must
be real positive and must be a decreasing function of time. The
best choice which yields positive decreasing temperature while
positive increasing entropy is obtained by setting $n=-1$ together
with $R_{0}\geq0$. In that case, typical behavior of temperature
and entropy are represented by the plots of Fig. \ref{fig4}.

\begin{figure}[h!]
 \begin{center}$
 \begin{array}{cccc}
\includegraphics[width=75 mm]{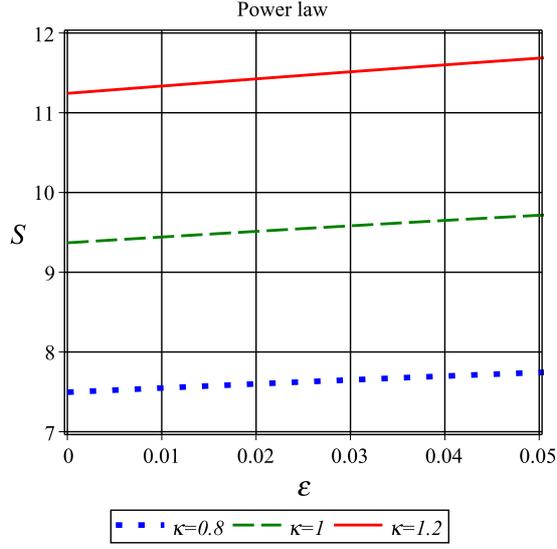}
 \end{array}$
 \end{center}
\caption{Entropy in terms of $\varepsilon$ for the power law model
for $R_{0}=1$, $n=-1$, $\kappa_{3}=0.2$, $\omega=-0.2$ and
$M=G=t=1$.}
 \label{fig5}
\end{figure}

We fix parameter $n$ where temperature is a decreasing function of
time with a positive real value. It is illustrated by the Fig.
\ref{fig4}(a). We can see that for the $n<-1$ temperature is
negative initially with possible asymptotic behavior, which
corresponds to a singular point, and is therefore considered as an
unphysical situation. Also, for the $n\geq0$, temperature is
increasing function of time. Hence $n=-1$ will be a judicious
choice for the power law model of $f(R, {\mathcal{L}})$ theory. We
can confirm the constraint on $n$ by analyzing entropy. Hence we
draw entropy in terms of time in the Fig. \ref{fig4}(b) and expect
that it is an increasing function of time to satisfy the second
law of thermodynamics. According to the Fig. \ref{fig4}(b) we can
see that $n\geq0$ gives us a maximum for the entropy, also it is
undefined at the late time perhaps due to some singularities. On
the other hand $n<-1$ yields another singularity and asymptotic
behavior. Hence, in order to have expanding universe we should
choose $n=-1\pm\epsilon$, where $\epsilon<1$ is an infinitesimally
small
constant.\\
Finally in Fig. \ref{fig5} we can see the effect of $\varepsilon$
on the entropy in power law model. We draw Fig. \ref{fig5} for the
case of $\omega=-0.2$ but it is seen that all cases of
$-\frac{1}{3}\leq\omega\leq0$ yields similar results. From the
above analysis it is quite natural to consider,
\begin{equation}\label{Ricci331}
R=\frac{R_{0}}{t}>0.
\end{equation}
which gives a well-defined thermodynamic relation. It is easy to
check that using the relation (\ref{sol3-2}) with $n=-1$ in the
Eq.(\ref{Ricci3}) yields similar behavior with the
Eq.(\ref{Ricci331}).

\begin{figure}[h!]
 \begin{center}$
 \begin{array}{cccc}
\includegraphics[width=75 mm]{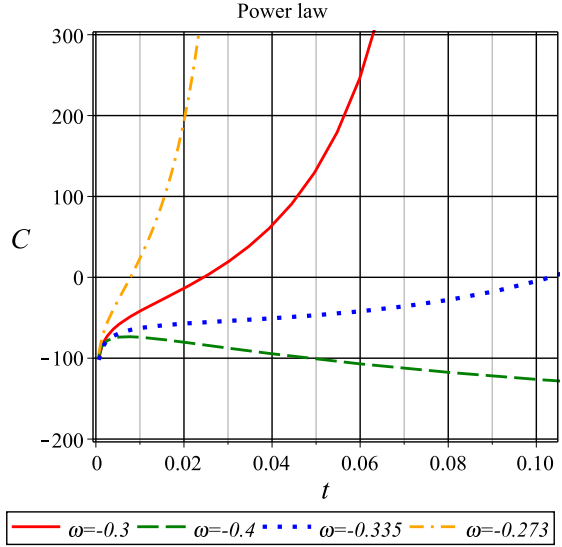}
 \end{array}$
 \end{center}
\caption{Specific heat in terms of $t$ for the power law model for
$R_{0}=1$, $\varepsilon=0.1$, $n=-1$ and
$M=\kappa_{3}=\kappa=G=1$.}
 \label{fig6}
\end{figure}

Then, using the relation (\ref{CV}) we can study specific heat and
thermodynamic stability of the model. It is illustrated by Fig.
\ref{fig6} which shows some instabilities of this model for
$\omega<-\frac{1}{3}$ exhibited by the green dashed line. We can
see that the whole trajectory lies in the negative region which is
undesirable. For $\omega>-\frac{1}{3}$ (yellow dot dashed line)
the model shows instability in the early phase, but eventually
settles to a stable configuration in the late universe. This is
because initially the trajectories lie in the negative region, but
they flip their signature and settle down in the positive region
at the late time. Also in the case of $\omega\approx-\frac{1}{3}$
(red line) there is some instability in the early universe which
goes to the stable phase at the late time. It may be fixed by
choosing $\epsilon\neq0$ (for example $n=-1.1$ or $n=-0.9$) and it
yields to a completely stable model. This quite clearly shows that
coupling baryonic matter with curvature produces models which are
thermodynamically favored compared to exotic matter coupling
models. Alternatively, we can study other models to find
thermodynamically stable system.

In the previous subsections we considered models with non-minimal
coupling between matter and curvature which was generated as a
result of series expansion of the functions. Now, we will consider
models involving non minimal coupling of the form
\begin{equation}\label{NMC2}
f(R,\mathcal{L})=\kappa_{4} f_{1}(R)+f_{2}(R)\mathcal{L},
\end{equation}
where $\kappa_{4}$ is a constant. We will study two separate
models of the above form namely the logarithmic model and
Starobinsky's model in the following subsections. In both these
models there will be a different $f(R)$ term with a NMC term added
to it. We name the models depending on the particular form of the
$f(R)$ term that we have considered.

\subsection{Logarithmic model}
\begin{figure}[h!]
 \begin{center}$
 \begin{array}{cccc}
\includegraphics[width=80 mm]{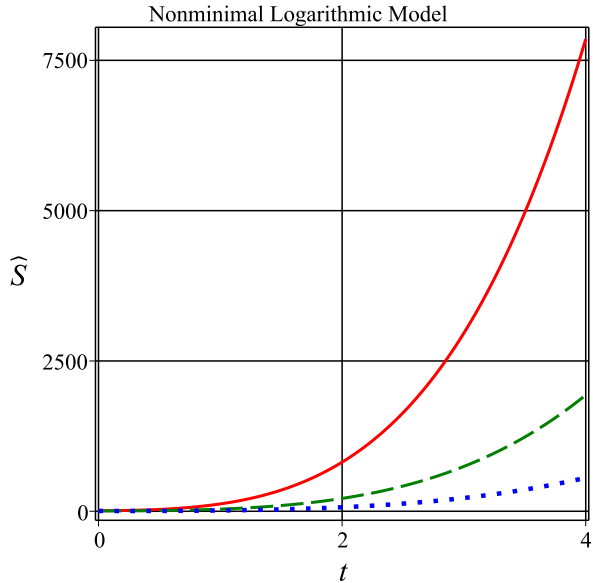}
 \end{array}$
 \end{center}
\caption{Modified entropy is plotted against time $t$ for the
logarithmic model. The initial conditions are taken as
$\mathcal{B}_{4}=c_{4}=a_{4}=b=G=1$. Blue dot for
$\mathcal{A}_{4}=1.2$, Green dash for $\mathcal{A}_{4}=1.4$, Red
solid for $\mathcal{A}_{4}=1.6$.}
 \label{fig7}
\end{figure}

In relation (\ref{NMC2}) we consider the following expressions
\begin{eqnarray}\label{Log}
f_{1}(R)&=&a_{4}\ln{bR}+c_{4}R,\nonumber\\
f_{2}(R)&=&\alpha R,
\end{eqnarray}
where $a_{4}$, $b$, $c_{4}$ and $\alpha$ are some constants. Here
$f_{1}(R)$ is taken in the logarithmic form and hence our
logarithmic model takes the shape of
\begin{equation}\label{NMC}
f(R,\mathcal{L})=\kappa_{4} (a_{4}\ln{bR}+c_{4}R)+\alpha R \mathcal{L},
\end{equation}
where $\alpha$ plays role of the coupling parameter between matter
and curvature. In case of $\alpha\neq0$, we realize non minimal
coupling. For $\kappa_{4}=c_{4}=1$ and $a_{4}=\alpha=0$ we recover
general relativity. Without loss of generality we can absorb
$\kappa_{4}$ in  $a_{4}$ and $c_{4}$. Then, by using the relations
(\ref{Lag0}) and (\ref{ah}) one can obtain,
\begin{equation}\label{Lag4-1}
{\mathcal{L}}=\frac{a_{4}r_{A}^{2}R^{2}(\ln{(bR)}-1)+6(c_{4}R^{2}-c_{4}r_{A}\dot{R}+6a_{4}R)}{6\alpha(3\alpha+2)R^{2}}.
\end{equation}
Hence, we can calculate entropy (\ref{S}), internal energy (\ref{E}) and specific heat (\ref{C}) for the logarithmic model. The specific heat (\ref{C}) written as,
\begin{equation}\label{C4-1}
C=\frac{\pi r_{A}^{2}\left(\dot{r}_{A}(\dot{r}_{A}-2)^{3}(\omega+\frac{1}{3}-\alpha)a_{4}r_{A}^{2}\ln{(\frac{6b(2-\dot{r}_{A})}{r_{A}^{2}})}+\cdots\right)}
{G({\dot{r}_{A}}^{2}-r_{A}\ddot{r}_{A}-2\dot{r}_{A})(\dot{r}_{A}-2)^{3}(\alpha+\frac{2}{3})},
\end{equation}
where we avoided writing long terms in the numerator. In order to
have $C\geq0$ both numerator and denominator should be positive.
Hence we find a possible solution from denominator and examine it
to obtain well defined solution. We find that by choosing
\begin{equation}\label{denominator4}
r_{A}=\mathcal{A}_{4}t+\mathcal{B}_{4},
\end{equation}
we can obtain positive specific heat and temperature which is a
decreasing function of time. $\mathcal{A}_{4}$ and
$\mathcal{B}_{4}$ are some constants which are constrained using
thermodynamic queries. For example, in order to have positive
temperature, specific heat and entropy we should choose positive
$\mathcal{A}_{4}$ and $\mathcal{B}_{4}$. By choosing unit values
for other parameter one can find that $\mathcal{A}_{4}>1.5$ is
necessary to have positive specific heat. Therefore, the entropy
(\ref{S}) using the solution (\ref{denominator4}) is expressed as
follows,
\begin{equation}\label{S4-1}
S=\frac{\pi\left[a_{4}(\mathcal{A}_{4}-2)(\mathcal{A}_{4}t+\mathcal{B}_{4})^{2}\ln{(\frac{6b(2-\mathcal{A}_{4})}{(\mathcal{A}_{4}t+\mathcal{B}_{4})^{2}})+\cdots}\right]}
{18G(\alpha+\frac{2}{3})(\mathcal{A}_{4}-2)(\mathcal{A}_{4}t+\mathcal{B}_{4})^{2}},
\end{equation}
where we avoid writing long terms in the numerator as before. One
can check that with any positive values of $\mathcal{A}_{4}$ and
$\mathcal{B}_{4}$ the entropy is decreasing function of time which
may be sign for violating the second law of thermodynamics. But it
is important to note that we should look at the modified entropy
(\ref{entropy-corrected}) to investigate the second law of
thermodynamics. In the Fig.\ref{fig7} we can see modified entropy
in terms of time for some values of constant $\mathcal{A}_{4}$. In
order to have completely increasing entropy we should choose
$\mathcal{A}_{4}>1.2$. Hence, by setting $\mathcal{A}_{4}=1.6$ we
have a completely well defined thermodynamical model where the
first and second laws of thermodynamics are satisfied and the
model is thermodynamically stable.

\begin{figure}[h!]
 \begin{center}$
 \begin{array}{cccc}
\includegraphics[width=80 mm]{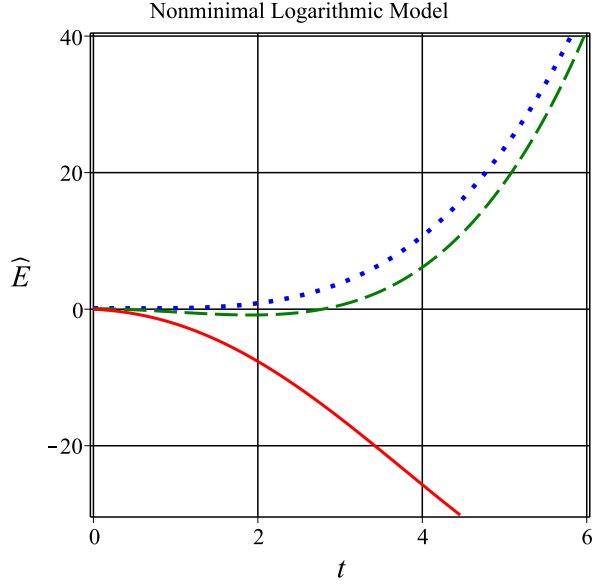}
 \end{array}$
 \end{center}
\caption{Internal energy is plotted against time $t$ for the
logarithmic model. The initial conditions are taken as
$\mathcal{B}_{4}=c_{4}=a_{4}=b=G=1$ and $\omega=-0.2$. Blue dot
for $\mathcal{A}_{4}=1.2$, Green dash for $\mathcal{A}_{4}=1.4$,
Red solid for $\mathcal{A}_{4}=1.6$.}
 \label{fig8}
\end{figure}

In the Fig. \ref{fig8} we draw internal energy for various values
of $\mathcal{A}_{4}$. It is initially negative in some
trajectories, which becomes positive at the late time. Hence there
is a minimum with the negative value for the internal energy. A
similar behavior for the Helmhuoltz free energy is found which
shows the stability of the model as confirmed by specific heat
analysis.

 \subsection{Starobinsky's model}
\begin{figure}[h!]
 \begin{center}$
 \begin{array}{cccc}
\includegraphics[width=60 mm]{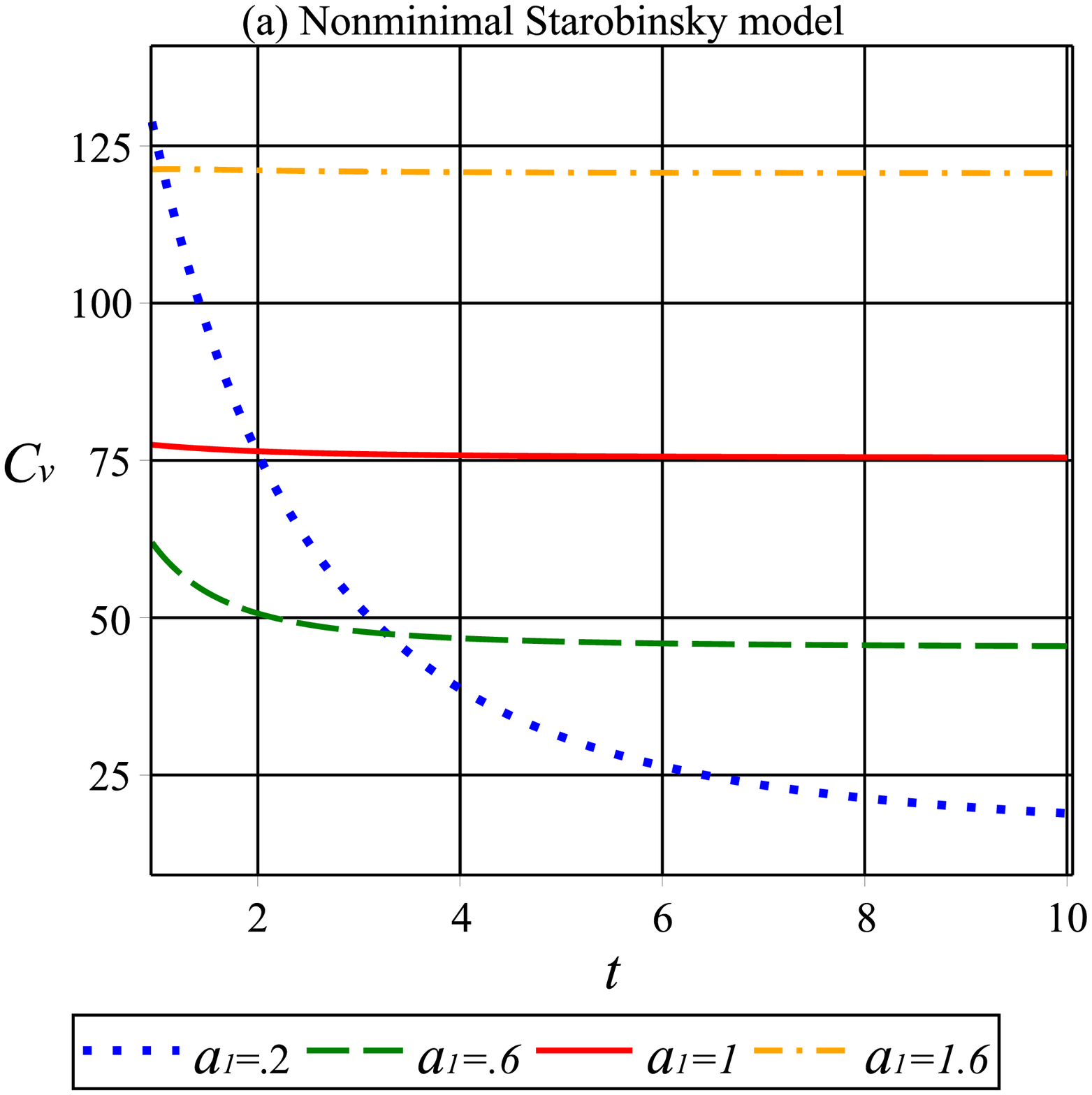}~~~~~~~\includegraphics[width=60 mm]{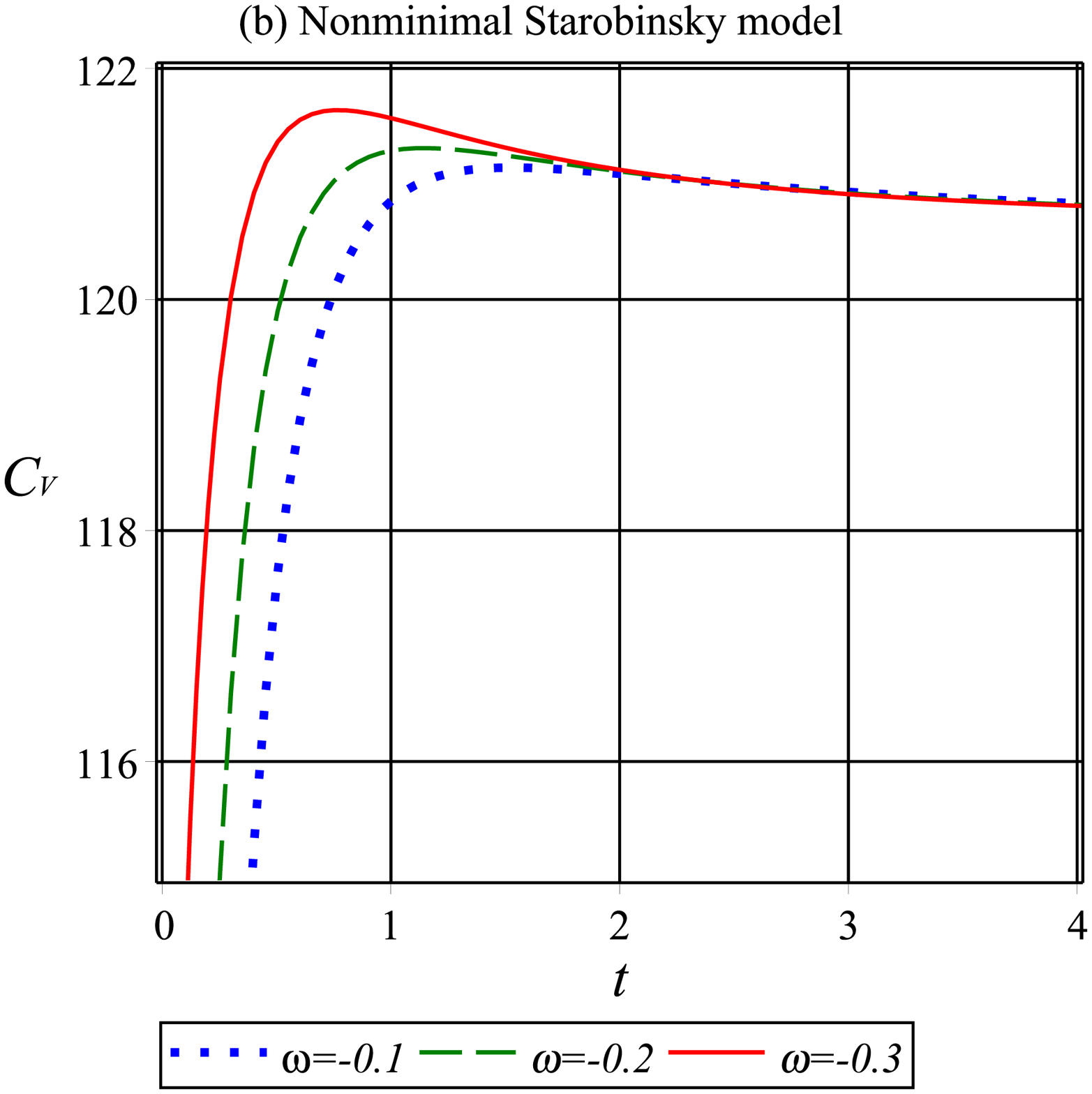}
 \end{array}$
 \end{center}
\caption{Specific heat at constant volume in terms of $t$ for the
NMC model with Starobinsky's form of $f(R)$ by choosing (a)
$\kappa_{5}=a=b_{1}=a_{0}=G=1$ and $\omega=-0.2$. (b)
$\kappa_{5}=a=b_{1}=a_{0}=G=1$ and $a_{1}=1.6$.}
 \label{CV4}
\end{figure}

Here we consider the $f_{1}(R)$ in the form of the famous
Starobinsky's model \cite{staro1, staro2} as given below
\begin{eqnarray}\label{Star}
f_{1}(R)&=&R+aR^{2},\nonumber\\
f_{2}(R)&=&b_{1}R,
\end{eqnarray}
where $a$ and $b_{1}$ are constants. This model is consistent with
the inflationary scenario of early universe. Studying this model
in a coupled form with matter will be very interesting. So, using
the above expressions in Eqn.(\ref{NMC2}) we get the ultimate
model as,
\begin{equation}\label{NMC-star}
f(R,\mathcal{L})=\kappa_{5} (R+aR^{2})+b_{1}R\mathcal{L},
\end{equation}
Here $b_{1}$ is the coupling constant which plays the role of a
controlling parameter. For $b_{1}\neq 0$ we get non-minimal
coupling.\\
Then, by using the relations (\ref{Lag0}) and (\ref{ah}) one can obtain,
\begin{equation}\label{Lag4-2}
{\mathcal{L}}=\frac{\kappa_{5}(6+12a(R+r_{A}\dot{R})-ar_{A}^{2}R^{2})}{6b_{1}(3\omega+2)}.
\end{equation}
Then, one can obtain,
\begin{equation}\label{S4-2}
S=\frac{3\pi\kappa_{5}}{G(3\omega+2)r_{A}^{4}}\left[\omega(r_{A}^{2}+24a)+16a-4r_{A}\ddot{r}_{A}a+2a\dot{r}_{A}(3\dot{r}_{A}-6\omega-10)+r_{A}^{2}\right].
\end{equation}
It yields the following expression for specific heat,
\begin{eqnarray}\label{C4-2}
C&=&\frac{24\pi\kappa_{5}ar_{A}^{2}\dddot{r}_{A}+3a(\omega+\frac{5}{3}-\dot{r}_{A})r_{A}\ddot{r}_{A}}{(3\omega+2)G(r_{A}\ddot{r}_{A}-\dot{r}_{A}^{2}+2\dot{r}_{A})}\nonumber\\
&-&\frac{3\dot{r}_{A}\left(a(\omega+\frac{5}{3})\dot{r}_{A}-\frac{a}{2}\dot{r}_{A}^{2}+\frac{1+\omega}{12}r_{A}^{2}-2a(\frac{2}{3}+\omega)\right)}{(3\omega+2)G(r_{A}\ddot{r}_{A}-\dot{r}_{A}^{2}+2\dot{r}_{A})}
\end{eqnarray}
As before by using the query that $C\geq0$ we choose,
\begin{equation}\label{Sol4-2}
r_{A}=a_{1}t+a_{0}.
\end{equation}
In that case we obtain,
\begin{equation}\label{T4-2}
T=\frac{2-a_{1}}{4\pi(a_{1}t+a_{0})}.
\end{equation}
Now, it is clear that $0<a_{1}<2$ and $a_{0}>0$ yields positive
temperature which is a decreasing function of time. Therefore, by
using the Eq.(\ref{entropy-corrected}) one can obtain,
\begin{eqnarray}\label{Corrected-entropy-4-2}
\hat{S}&=&\frac{72\kappa_{5}\pi\left[\frac{a_{1}^{2}}{4}(a+\frac{1}{6}(1+\omega)t^{2})-a_{1}(\omega+\frac{2}{3})a(a_{1}t+a_{0})^{4}\ln{(a_{1}t+a_{0})}\right]}{G(3\omega+2)(a_{1}t+a_{0})^4}\nonumber\\
&-&\frac{72\kappa_{5}\pi\left[((\omega+\frac{5}{3})\frac{a}{2}-\frac{a_{0}(1+\omega)t}{12})a_{1}-(\omega+\frac{2}{3})a-\frac{a_{0}^{2}}{24}(1+\omega)\right]}
{G(3\omega+2)(a_{1}t+a_{0})^4}.
\end{eqnarray}
Therefore, by using the Eq.(\ref{CV}) one can obtain,
\begin{equation}\label{CV4-2}
C_{V}=\frac{72\pi \kappa_{5}}{G(3\omega+2)(a_{1}t+a_{0})^4}\left(aa_{1}^{3}(\omega+\frac{2}{3})t^2(a_{1}^{2}t^{2}+4a_{0}a_{1}t+6a_{0}^{2})+\mathcal{C}\right),
\end{equation}
where
\begin{eqnarray}\label{CVC4-2}
\mathcal{C}&\equiv&\left((1+4(\omega+\frac{2}{3})a_{0}^{3}t)a+\frac{1+\omega}{12}t^2\right)a_{1}^{2}\nonumber\\
&+&\left(((\omega+\frac{2}{3})a_{0}^{4}-2\omega-\frac{10}{3})a+\frac{a_0(1+\omega)t}{6}\right)a_{1}\nonumber\\
&+&4(\omega+\frac{2}{3})a+\frac{\pi(1+\omega)a_{0}^{2}}{12}.
\end{eqnarray}
We find from the first law of thermodynamics that this model
yields positive specific heat as illustrated by Fig. \ref{CV4}
(a). In the Fig. \ref{CV4} (b) we show there are some special
cases that heat capacity rises to a maximum, which is known as a
Schottky anomaly where the system attains such a temperature where
there is a possibility of thermally excited transitions between
two states of the system, however it is different with phase
transition (asymptotic behavior of specific heat).

\begin{figure}[h!]
 \begin{center}$
 \begin{array}{cccc}
\includegraphics[width=80 mm]{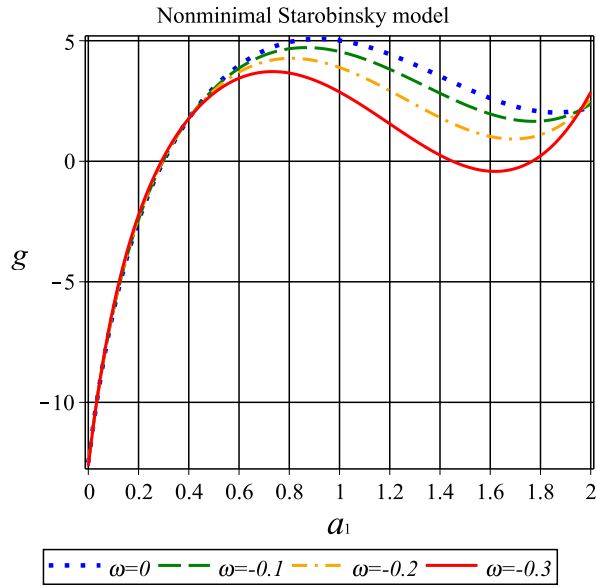}
 \end{array}$
 \end{center}
\caption{Gibbs free energy in terms of $a_{1}$ for the NMC model
with Starobinsky's form of $f(R)$ by choosing
$\kappa_{5}=a=b_{1}=a_{0}=G=t=1$. The Gibbs free energy is a
totally decreasing function in some range of the domain of
$a_{1}$.}
 \label{fig9}
\end{figure}

In order to confirm stability of the model we can study Helmholtz
free energy (F) and Gibbs free energy $g$. By suitable choice of
model parameters we can have $F$ with a minimum value and $g$ as a
decreasing function of time. In the Fig. \ref{fig9} we represent
the behavior of Gibbs free energy in terms of $a_{1}$ and see that
by suitable choice of model parameters the model can be made
stable (Gibbs free energy is totally decreasing function). It
gives us further constraints on parameter $a_{1}$ to have upper
and lower bounds as $0.8\leq a_{1}\leq1.7$ ($g$ is decreasing in
this range) for selected values of other parameters. It means that
we can have a completely stable model in that range of $a_{1}$.
This monotonic nature of $g$ is a unique feature of the
Starobinsky's model.

\section{Discussions and Conclusions}
Here, we have investigated the thermodynamic properties of
$f(R,\mathcal{L})$ gravity, where spacetime curvature $R$ is
coupled with baryonic matter $\mathcal{L}$. Models involving both
minimal and non-minimal coupling have been studied. In our
investigation we considered thermodynamic parameters like
temperature of the apparent horizon, entropy, specific heat at
constant volume, total internal energy, enthalpy, Gibbs free
energy and Helmholtz free energy. Depending on the results
obtained we can comment on the stability of the model from the
thermodynamic point of view. In order to have a well-defined model
it should support an increasing entropy, non-negative temperature,
non-negative specific heat, a decreasing Gibbs free energy and a
minimum value of Helmholtz free energy. These characterize a
realistic cosmological model.

We studied various models of $f(R,\mathcal{L})$ theory. The first
model studied was general relativity, where the coupling between
$R$ and $\mathcal{L}$ is minimal in nature. The study showed that
the model is consistent with the stability requirements which is
quite obvious. Then, we studied four different models involving
non-minimal coupling between matter and curvature. The first one
of these was the exponential model, which generates non-minimal
terms on series expansion. From the study it was seen that the
apparent horizon could be given in the form of a hyperbolic
tangent. It was also seen that this model may lead to negative
$C_V$ at late times which is not physical. Next we considered the
power law model, which again yields non-minimal terms on
expansion. For this model we considered power law form of
curvature and computed various thermodynamic parameters for
positive curvature. The trend of temperature $T$ and entropy $S$
with respect to time was investigated for this model for various
values of the power law parameter $n$. From these results, the
parameter $n$ was constrained to $n=-1$ to give realistic
thermodynamically stable models. Plots of $S$ vs $\varepsilon$ and
$C$ vs $t$ were also generated to study the model characteristics.
It was seen that the baryonic matter couplings are much more
favored thermodynamically compared to exotic matter couplings for
this model. Logarithmic model was studied where a logarithmic form
of $f(R)$ was considered with a non-minimal term. Various
parameters were studied to check the thermodynamical viability of
the model. Plots for entropy and internal energy was generated for
various values of the parameter $\mathcal{A}_{4}$. The model was
found to be fairly stable thermodynamically. Finally we studied an
NMC model with Starobinsky's form of $f(R)$. Here the trend of
Gibbs free energy $g$ was checked with respect to time $t$. It was
seen that $g$ was a totally decreasing function of time in some
range of the domain, thus indicating the relative stability of the
model. Monotonic nature of $g$ allows us to constrain the model
parameters significantly. Other thermodynamic parameters were
investigated under this model and it was seen that there was ample
scope to further constrain the model parameters for suitable
initial conditions.

This study gives us a detailed thermodynamic prescription of the
recently proposed $f(R,\mathcal{L})$ theories. Since we have
considered different types of couplings between matter and
curvature and kept the models as generic as possible the span of
the work covers a large class of $f(R,\mathcal{L})$ theories. It
is hoped that this work will considerably develop our
understanding of $f(R,\mathcal{L})$ theories and enrich the
existing literature on the topic. Cosmological viability of these
models will be very important because of its nonexotic nature of
the matter content as discussed earlier. Therefore a study on the
various cosmological aspects of this theory will be very
interesting and will be attempted in a future work. Moreover
comparing our parameter constraints from this thermodynamical
study with an observational data analysis mechanism will also be
an interesting proposition.

\section*{Acknowledgments}
B.P. would like to thank Iran Science Elites Federation. P.R.
acknowledges Inter University Centre for Astronomy and
Astrophysics (IUCAA), Pune, India, for awarding Visiting
Associateship.

\end{document}